# Magnetic Measurements and Alignment Results of LQXFA/B Cold Mass Assemblies at Fermilab

J. DiMarco, P. Akella, G. Ambrosio, M. Baldini, G. Chlachidze, S. Feher, J. Nogiec, V. Nikolic, S. Stoynev, T. Strauss, M. Tartaglia, P. Thompson, D. Walbridge

*Abstract*— MQXFA production series quadrupole magnets are being built for the Hi-Lumi (HL) LHC upgrade by the US Accelerator Upgrade Project (US-HL-LHC AUP). These magnets are being placed in pairs, as a cold mass, within cryostats at Fermilab, and are being tested to assess alignment and magnetic performance at Fermilab's horizontal test stand facility. The ~10 m - long assembly must meet stringent specifications for quadrupole strength and harmonic field integrals determination, magnetic axis position, and for magnet variations in positioning and local field profile. This paper describes the results of the magnetic and alignment measurements which characterize the first LQXFA/B assembly.

## I. INTRODUCTION

The US-HiLumi Accelerator Upgrade Project (AUP), comprised of Fermilab, Brookhaven National Lab (BNL), and Lawrence Berkeley National Lab (LBNL), are currently in the process of producing 20 4.5 m-long $Nb_3Sn$ superconducting magnets for the final-focus interaction regions of the LHC. The final step in this production, after assembling pairs of magnets into a single cold-mass and inserting them into a cryogenic vessel, is horizontal testing at nominal LHC temperature and field conditions to demonstrate compliance with performance specifications in their final configuration.

The first article cryo-assembly (CA) in the series, LQXFA/B01, containing magnets MQXFA03 (referred herein as A03) and MQXFA04 (A04), has recently undergone testing at Fermilab's horizontal magnet test facility. An overview of fabrication, facilities, and quench performance test results for this CA is given elsewhere [1][2][3]. Here we present the results of the LQXFA/B01 magnetic measurements, including alignment, integrated quadrupole strength, geometrical field harmonics, and local magnet variations, as well as comparison to requirements specifications.

Submitted for review September 19, 2023

This work was supported by the U.S. Department of Energy, Office of Science, Office of High Energy Physics, through the US HL-LHC Accelerator Upgrade Project..

The authors are with Fermi National Accelerator Laboratory, P.O. Box 500, Batavia, IL 60510, USA, (corresponding author e-mail: dimarco@fnal.gov).

Color versions of one or more of the figures in this article are available online at http://ieeexplore.ieee.org

## II. FIELD DEFINITION AND MEASUREMENT SYSTEM

The measured field of a quadrupole is expressed in a standard form of harmonic coefficients defined in a series expansion of a complex field function

$$B_y + iB_x = B_2\, 10^{-4} \sum_{n=1}^{\infty} (b_n + ia_n)\left(\frac{x+iy}{R_{ref}}\right)^{n-1}$$

where $B_x$ and $B_y$ are the field components in Cartesian coordinates, $b_n$ and $a_n$ are the normal and skew coefficients at reference radius $R_{ref}$, normalized to the main field, $B_2$, and scaled by a factor $10^4$ so as to report the harmonics in convenient 'units'. All measurements utilize $R_{ref}$ of 50 mm, and harmonics are reported after feed-down correction for the probe being off-center. Though the magnets are assembled 'back-to-back' in the cryostat and use a 'thru-bus' to power them so they function as a single unit (Fig. 1), harmonics are reported for each magnet individually in a frame that views each magnet from its lead-end, and powers the magnet to have a negative normal quadrupole as main field. For these individual magnet results, a right-handed measurement coordinate system is defined with the origin at the center of the magnetic aperture, centered axially on the magnetic length, with x-axis to the right, and z pointing towards the lead end. This matches the reporting frame requested by CERN [4], and also the frame used for harmonics results during fabrication obtained at LBNL [5] and vertical magnet testing at BNL [6].

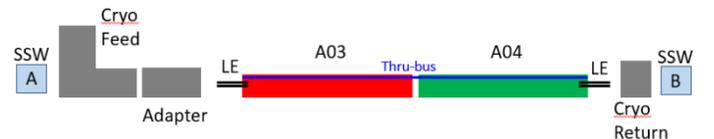

**Fig. 1.** Schematic of test stand showing cryogenic com-ponents, SSW stages, and lead ends (LE) of magnets A03/A04

Magnetic measurements are performed with the systems and techniques described in [7], principally a Single Stretched Wire system (SSW), and translated rotating coil (RC) probe fitted with laser tracker targets. The rotating coil is made-up of 3 PCB-based probes with dipole and quadrupole buck windings: two back-to-back probes of 109 mm length (matching the transposition pitch of the cable in the magnet), and one of 436 mm length. The probe has a local encoder and 22 m-long driveshaft/push-tube which connects to a stepper motor external to the magnet on a drive rail [7]. The test plan

 

included an axial scan at nominal operating current (16233 A) with steps of the short, 109 mm-long, probes of the RC. However, cryogenic operational issues limited the duration of high-field testing, and consequently, rotating coil measurements at nominal current were made with the 436 mm step of the long probe on the RC. This helped reduce the length of scan time, and thus to avoid a leads quench caused by temperature excursion. Two scans were made, each completing in about 30 minutes. In order to have the measurements necessary to determine local field variation, a fine step axial scan ('Z-scan') was also performed at 6000 A.

### III. MEASUREMENT RESULTS AND DISCUSSION

*A. Integral Strength and Transfer Function*

The integral gradient strength, $gdl$, of LQXFA/B01 was measured with both rotating coil and SSW at nominal operating current at 1.9 K. The integral start point position was determined from the fine step 6000 A Z-scan so as to include the full length of the CA and be symmetric with respect to the end regions. The actual probe increment could vary from the position requested (Fig. 2). However, since the probe position is measured accurately by the laser tracker, the integral was calculated using the average body field value ($g_{body_{ave}}$) over the positions bracketing the body length ($L_{body}$) and including the end contributions (Lead End and Non-Lead End). That is,

$$\int gdl = \int LE + \int NLE + g_{body\_ave} L_{body} \quad .$$

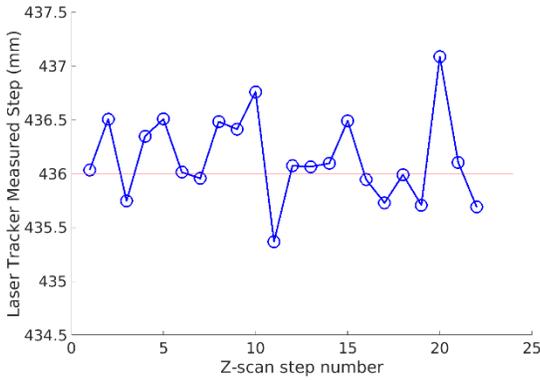

**Fig. 2.** Actual distance for nominal 436 mm probe step.

Current was measured by Holec transductor and Agilent 3458 DVM. The probe calibration is based on the leading short probe, the PCB of which was fabricated with LDI (Laser Direct Imaging) and which has dipole-quad bucking ratio (i.e. ratio of quad field before/after suppression) of almost 700, indicating a good level of precision in fabrication. This should provide accuracy better than ~0.15% and will be further calibrated against SSW to confirm/improve this. The integral results at 16233A are summarized in Table I showing total integrated field, magnetic length, and transfer function in the body. Agreement between integral strength of the two scans is at the 1 unit level. The CA integral of 1119.65 T at this current is ~0.5% above the acceptance requirement of 1114 T.

The SSW integral strength result was determined by extrapolating measured strength as a function of reciprocals of the wire tensions (reciprocal squared frequencies) to effectively remove linear wire susceptibility effects as shown in Fig. 3 [8]. Using the frequency is in principle better than tension, since the tension is measured by a gauge which can have additional mechanical contributions. This integral gives a value of 1119.1 T, in good agreement with the rotating coil, but this value has a rather high uncertainty (6 T) relative to requirements. The uncertainty seems to stem from the frequency determination, which was only done at the level of 1%; efforts will be made to improve this for the next CA.

TABLE I
LQXFA/B01 ROTATING COIL STRENGTH SUMMARY

| Cryo-Assembly Magnet: | A04 | A03 |
|---|---|---|
| Integral Gdl (T): | 559.95 | 559.70 |
| Magnetic length (m): | 4.213 | 4.216 |
| Body field TF (T/m/kA): | 8.187 | 8.178 |
| Magnet center separation (m): | 4.7721 | |

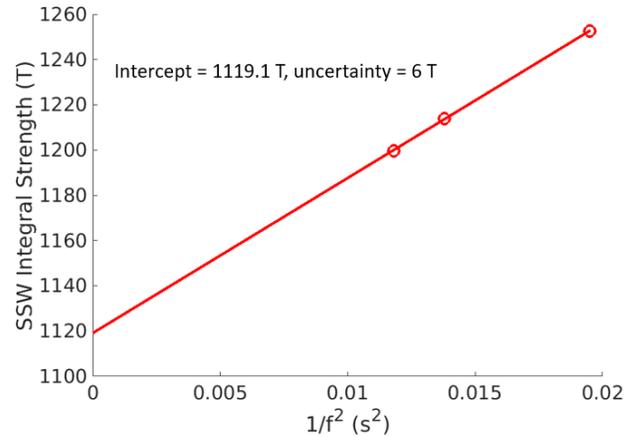

**Fig. 3.** SSW integral strength measurement.

*B. Magnet Alignment*

SSW alignment measurements were performed at 1.9 K with 10 A, 7.8125 Hz AC current, during the second test cycle. The magnets were powered individually, and sag effects of the 18.3 m-long, 100 μm diameter BeCu wire were compensated by performing calibration as a function of wire tension or wire vibration frequency [9] up to a maximum tension of ~ 850 g. The asymmetric axial positioning of the magnets with respect to wire ends is compensated for as part of the measurement technique [9]. Note that magnetization effects of these low current measurements do not impact the alignment, as the effect of harmonics is small (especially with wire motions being less than one third of the reference radius), and any main field effects are quadrupole symmetric [10]. The alignment places the wire on the average center of both magnets, and its location is then transferred to magnet fiducials via survey with laser tracker. Relative magnet roll angle between the two magnets was 2.3 mrad and within acceptance criteria. The alignment summary is shown in Fig 4 and Table II. The deviation of the magnet ends from the average-axis line was generally within the +/- 0.5 mm of the acceptance criteria, except for A04 which had vertical offsets that exceeded this by about 0.2 mm.

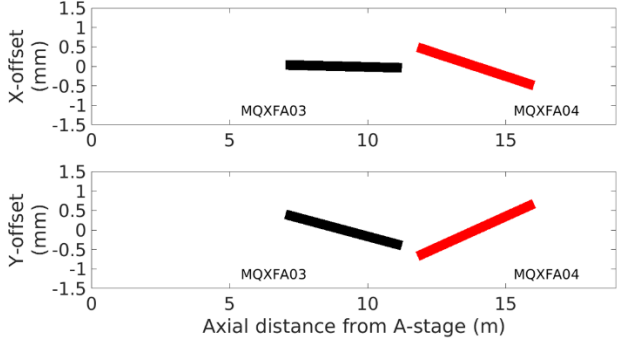

**Fig. 4.** SSW alignment measurements, 1.9K

TABLE II
LQXFA/B01 ALIGNMENT SUMMARY (IN MM)

| Offset | X | Y |
|---|---|---|
| MQXFA03 Lead End: | 0.042 | 0.394 |
| MQXFA03 Interface End: | -0.030 | -0.402 |
| MQXFA04 Interface End: | 0.498 | -0.676 |
| MQXFA04 Lead End: | -0.482 | 0.681 |

C. Integral Harmonics

Integral harmonics are measured at nominal current at 1.9 K during the rotating coil scans which also measure the strength integrals described previously. Rotation rate is 4 Hz. The rotating coil PCBs have been calibrated for radial and transverse offsets in mounting at the level of a few microns using an in-situ technique [11]. The calibration is also anchored to the leading short coil, as it was for strength. The normalized integral harmonics are calculated over the Z-positions after rotation so that each $A_2 = 0$, according to

$$b_{n_{int}} + ia_{n_{int}} = \frac{\sum_{Z=1}^{Z_{tot}} B_{n_z} + iA_{n_z}}{\sum_{Z=1}^{Z_{tot}} |B_2|_z} 10^4$$

These are shown in comparison to tolerances in Fig 5.

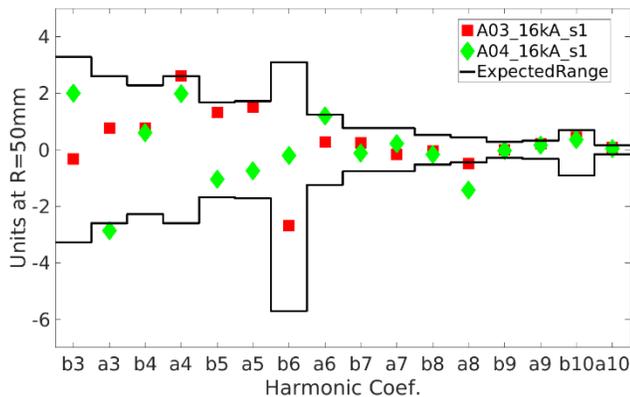

**Fig. 5.** Integral harmonics at nominal operating current (16233A) compared to 3σ range + uncertainty.

The harmonics fit well within the range brackets for both magnets of the CA except for $a_3$, and more so, $a_8$, of magnet

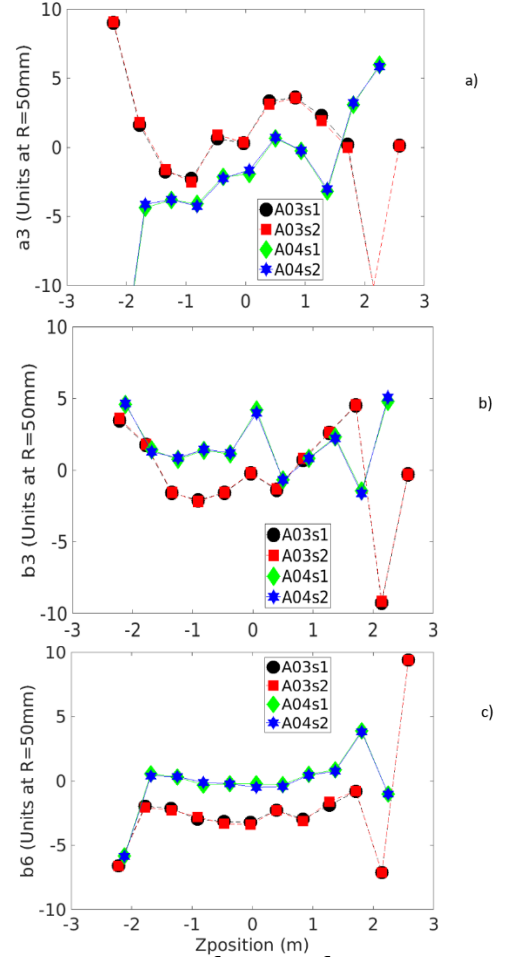

**Fig. 6.** Harmonics a) $a_3$ b) $b_3$ and c) $b_6$ as a function of axial position at nominal current, 1.9 K, for scans S1 and S2 of magnets A03 and A04.

A04. These large values were observed in previous measurements and are suspected to stem from the magnet having been fabricated with one coil of different shim layout than the others [5]. Plots of the individual harmonics which compromise these averages are given as a function of axial position for a sample of harmonics in Fig 6, with repeatability better than 0.1 units seen between the two scans on each magnet. The values of integral harmonics are also summarized in Table III.

TABLE III
LQXFA/B01 ROTATING COIL HARMONICS SUMMARY

| | A04 | A03 | A04 | A03 |
|---|---|---|---|---|
| n | $b_n$ | $b_n$ | $a_n$ | $a_n$ |
| 3 | 2.00 | -0.33 | -2.87 | 0.75 |
| 4 | 0.59 | 0.76 | 1.98 | 2.61 |
| 5 | -1.04 | 1.32 | -0.75 | 1.51 |
| 6 | -0.20 | -2.69 | 1.19 | 0.26 |
| 7 | -0.11 | 0.25 | 0.21 | -0.17 |
| 8 | -0.16 | -0.06 | -1.43 | -0.49 |
| 9 | -0.04 | -0.01 | 0.16 | 0.21 |
| 10 | 0.35 | 0.45 | 0.04 | 0.08 |

## D. Local Angle Variations

The local angular twist variations (or, equally, local skew quadrupole) of the magnet assembly was measured at 6000 A, 4.2 K using the rotating coil, with 109 mm scan step corresponding to the length of the each of the dual short probes. The lower current helped minimize the likelihood of a quench caused by a temperature excursion in the cryogenics during the fairly lengthy measurement (about 2 hours), but still represents the variations expected at the nominal current. For each Z position, the trailing probe provides a relative orientation of the angle measurement of the lead probe. Combining these, as described in [7], and using a calibration of the offset angle between the two probes (7.05 mrad), which would otherwise mimic an overall twist, the local variation at each point along the magnets is determined. The results are shown in Fig 7 and are well within the +/-2 mrad variations of the acceptance criteria. These also closely resemble the variations seen warm during magnet fabrication [5].

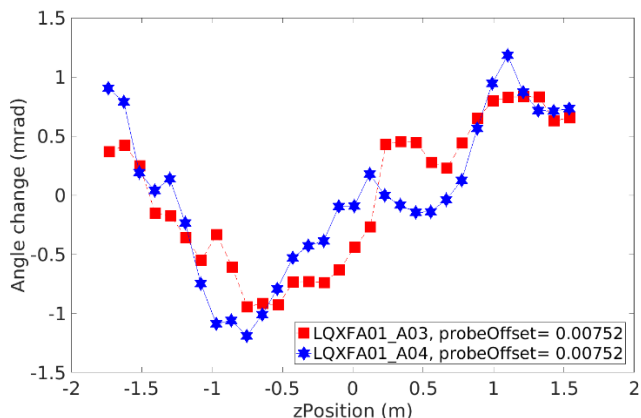

**Fig. 7.** Local angular twist along the magnets.

## E. Local Magnet Center Offset

As the rotating coil follows the anti-cryostat (or Warm Bore Tube (WBT)), it may be offset transversely with respect to the magnetic field - this can be measured directly by the probe from the harmonic feed-down of quadrupole creating apparent dipole field. To distinguish these probe position changes from field changes caused by local variations in the magnet structure itself, a laser tracker (LT) is used to measure the mechanical offset of the probe during the scans. Removing the LT measured mechanical offset from the probe magnetic offset will give the local quadrupole center variation caused by the magnet. Accomplishing this in our measurements required calibration to locate the probe rotation axis (at the axial position of each probe) wrt the LT targets. The calibration consisted of mounting targets on the PCBs and monitoring them during probe rotation to determine axis. A Helmert transformation is then used to determine and append the mechanical probe center as a data point for each axial position, therefore making a proper projection, including angular corrections across the distance from the targets at the end of the probe to the probes themselves. The pitch and yaw angles of the probe during Z-scan were as high as ~ 2 mrad, which over the distance from LT targets to probe (~0.5 m) implies corrections at the level of 1mm, and so significant for the level of accuracy needed.

The LT coordinate system can, if desired, be transformed to lie along the average two-magnet axis by referencing fiducials in the building common to both the SSW survey and RC measurements LT data. For our purposes, the difference between the RC offset and mechanical LT offset was taken with the axial LT coordinate defined between centroids at the beginning and end of the beam pipe. This was then adjusted to be on the individual magnet axes as defined by the average variations, so that we could focus on those local variations. The residual magnetic offsets for LQXFA/B01 are shown on Fig 8 and are well within the 0.5 mm criteria. These are also consistent with the fabrication measurements at LBL [5].

Note that in principle, a roll transformation of the LT data (taken with respect to gravity) and the average SSW magnet roll angle should be made, since the probe measures wrt the quadrupole field, which has an overall tilt on the test stand wrt gravity. But since the offsets of the probe center were small (2-3 mm) and the average roll on the test stand was about 3.4 mrad, the sine projections are negligible, on the order of 10 μm and so are not included.

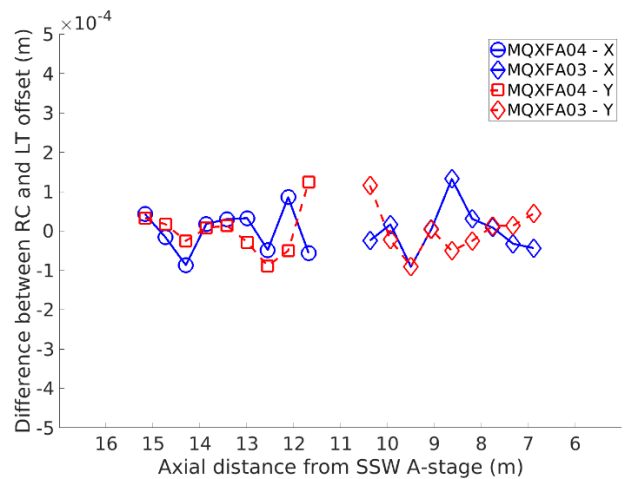

**Fig. 8.** Local magnet center offset variation.

## V. CONCLUSION

Magnetic measurements on the first LQXFA/B cold mass assembly were able to determine all quantities of interest, with precision adequate to characterize the magnet. The magnetic parameters are within the criteria, except for the $a_3$ (marginally) and $a_8$ harmonics of magnet A04, and 0.2 mm alignment offset at the ends of magnet A04 with respect to the average axis of the two magnets.

ACKNOWLEDGMENT

The authors gratefully acknowledge the contributions and dedication to these efforts by the technical staff of Fermilab's Applied Physics and Superconducting Technology Division.

## REFERENCES

[1] S. Feher, et al., "AUP first pre-series cryo-assembly design production and test overview", IEEE Trans. Appl. Supercond., to be published.

[2] G. Chlachidze, et al.,"Fermilab's horizontal test stand upgrade overview and commissioning", IEEE Trans. Appl. Supercond., to be published.

[3] M. Baldini, et al., "Quench performance of the First Pre-series AUP Cryo-Assembly", IEEE Trans. Appl. Supercond., to be published.

[4] R. Carcagno, G. Sabbi and P. Ferracin, "Acceptance criteria Part A: MQXFA magnet", no. LHC-MQXFA-ES-0004, Oct. 2018.

[5] Wang, X., et al., "Field Quality of the 4.5-m-Long MQXFA Pre-Series Magnets for the HL-LHC Upgrade as Observed During Magnet Assembly", IEEE Trans. Appl. Supercond. 32 (2022) 4002405, DOI 10.1109/tasc.2022.3156540.

[6] A. Ben Yahia, "Training performance and magnetic measurements during vertical testing of the MQXFA magnets for HL-LHC", IEEE Trans. Appl. Supercond., to be published.

[7] DiMarco, J., et al. "Magnetic measurements of HL-LHC AUP cryo-assemblies at Fermilab." IEEE Transactions on Applied Superconductivity 32.6 (2022): 1-7.

[8] N. Smirnov, et al., "Focusing Strength Measurements of the Main Quadrupoles for the LHC", IEEE Transactions on Applied Superconductivity 16(2):261 – 264, DOI:10.1109/TASC.2006.871218

[9] J. DiMarco, et al., "Alignment of production quadrupole magnets for the LHC interaction regions", July 2003, IEEE Transactions on Applied Superconductivity 13(2):1325 – 1328, DOI:10.1109/TASC.2003.812659

[10] [2] J. DiMarco, J. Krzywinski, "MTF Single Stretched Wire System", MTF-96-001, 1996.

[11] J. DiMarco, G. Severino, P. Arpaia, "Calibration technique for rotating PCB coil magnetic field sensors", Sensors and Actuators A: Physical, Volume 288, 2019, Pages 182-193, ISSN 0924-4247, https://doi.org/10.1016/j.sna.2019.02.014